# The Influence of Longitudinal Space Charge Fields on the Modulation Process of Coherent Electron Cooling

G. Wang[1,*], M. Blaskiewicz[1], V. N. Litvinenko[1,2]

[1]*Collider-Accelerator Department, Brookhaven National Laboratory, Upton, New York 11973-5000, USA*

[2]*Department of Physics & Astronomy, Stony Brook University, Stony Brook, New York 11794-3800, USA*

The initial modulation in the scheme for Coherent electron Cooling (CeC) rests on the screening of the ion charge by electrons. However, in a CeC system with a bunched electron beam, inevitably, a long-range longitudinal space charge force is introduced. For a relatively dense electron beam, its force can be comparable to, or even greater than the attractive force from the ions. Hence, the influence of the space charge field on the modulation process could be important. If the 3-D Debye lengths are much smaller than the extension of the electron bunch, the modulation induced by the ion happens locally. Then, in that case, we can approximate the long-range longitudinal space charge field as a uniform electric field across the region. As detailed in this paper, we developed an analytical model to study the dynamics of ion shielding in the presence of a uniform electric field. We solved the coupled Vlasov-Poisson equation system for an infinite anisotropic electron plasma, and estimated the influences of the longitudinal space charge field to the modulation process for the experimental proof of the CeC principle at RHIC.

## I. INTRODUCTION

Idea of Coherent electron Cooling (CeC) first was introduced by Y. Derbenev in 1980s [1]. In 2007, V. N. Litvinenko and Y. Derbenev developed detailed theory of the Free Electron Laser (FEL) based CeC scheme [2, 3], wherein FEL is used as an amplifier, and the time-of-flight dependence upon hadrons' energy is exploited to cool them. Estimates suggest that this scheme potentially could cool a high-energy, high-intensity ion beam in modern hadron accelerators, such as the RHIC, LHC and the proposed eRHIC. More recently, a similar CeC concept based upon amplification via micro-bunch instability was proposed by Ratner [4], which he called Micro-bunched Electron Cooling (MBEC). This technique potentially has a much larger bandwidth than other CeC schemes. A CeC system comprises of three sections: Modulator, amplifier, and kicker. In the modulator, the ion beam and the electron beam are merged. Each ion creates an electron-density modulation around itself through process shielding, or screening. The modulation in electron density then is amplified in the CeC amplifier and acts back on the ion in the kicker section, so reducing the error in the ions energy. By coupling the transverse- and longitudinal-motions [2, 3], the oscillations in all three degrees of freedom are cooled.

The modulation of CeC relies on the Coulomb interaction between electrons and ions. The dynamics of this process in a uniform anisotropic electron plasma was investigated previously and it is shown that, at the limit of the cold electron-beam, the results obtained are reduced to those from the hydro-dynamical model [3, 5]. In those calculations, it is assumed that electrons have uniform spatial distribution and, hence, that there is no net space charge field in the un-perturbed electron plasma. This assumption is valid if the spatial extension of the electron bunch is much larger than the Debye lengths in all three dimensions, and the ion is located close to the center of the electron bunch. However, for an electron bunch with high density and an ion interacting with electrons at a distance away from the bunch's center, the electrons surrounding the ion may see a net longitudinal space charge force comparable to, or even greater than the attractive force from the ion. This makes it necessary to account for the long-range space charge field while analyzing the process of modulation.

---

*gawang@bnl.gov

In this work, we do not assume that the net long-range space charge field is negligible, whilst still assuming that the spatial distribution of electrons is smooth and its spatial extension is much larger than the Debye lengths in all three dimensions. With these considerations, it still is possible to make the approximation that the electrons participating in shielding a specific ion have a uniform spatial distribution. Also, the long-range space charge field perceived by these electrons can be considered as uniform. Consequently, the modulation process can be described by the self-consistent Vlasov-Poisson equation system for uniform electron plasma in the presence of a moving ion and an external electric field. By linearizing the Vlasov equation, we can solve analytically the equation system for the κ-2 velocity distribution, and so obtain the density modulation in the simple form of 1-D integral.

The paper is organized as follows. In section II, we derive the linearized Vlasov-Poisson equation for the system, and solve the equations for the background electrons. The linearized Vlasov-Poisson equation system is solved in section III, so yielding the electron-density modulation induced by the moving ion. In section IV, we give several numerical examples of the influence of the longitudinal long-range space charge field on the modulation process in the proof-of-principle experiment of CeC and in the proposed CeC systems for the eRHIC and LHC. We assess the reduction of the longitudinal space charge field due to screening by the beam pipe in section V. Section VI presents the summary.

## II. LINEARIZED VLASOV-POISSON EQUATIONS

It is convenient to choose the reference frame as the rest frame of the ion, wherein the velocities of electrons are non-relativistic. Such motion of the particles in this frame allows us to use electrostatic Poisson equation as a good approximation for the evolution of the electric fields.

Let $f(\vec{x},\vec{v},t)$ be the distribution of the electrons' phase space density at time $t$ with the initial distribution at $t=0$ of

$$f(\vec{x},\vec{v},0) = f_0(\vec{v}). \tag{1}$$

For $t > 0$, the phase space distribution function is determined by the coupled Vlasov-Poisson equation system:

$$\frac{\partial}{\partial t} f(\vec{x},\vec{v},t) + \vec{v} \cdot \frac{\partial}{\partial \vec{x}} f(\vec{x},\vec{v},t) - \frac{e}{m_e} \left[ \vec{E}_{ext} - \vec{\nabla} \phi_{ind}(\vec{x},t) \right] \cdot \frac{\partial}{\partial \vec{v}} f(\vec{x},\vec{v},t) = 0 , \quad (2)$$

and,

$$\nabla^2 \phi_{ind}(\vec{x},t) = -\frac{1}{\varepsilon_0} \left[ Z_i e \delta(\vec{x}) - e \int_{-\infty}^{\infty} f(\vec{x},\vec{v},t) d^3v \right], \quad (3)$$

where $\vec{E}_{ext}$ is the uniform space charge field at the location of the ion, and $Z_i e$ is the ion's electric charge. The electric potential, $\phi_{ind}(\vec{x},t)$, is induced both by the ion and the electrons' response to the ion's field. To linearize eq. (2), we write the phase space density of electrons as

$$f(\vec{x},\vec{v},t) = f_{BG}(\vec{x},\vec{v},t) + f_{ind}(\vec{x},\vec{v},t), \quad (4)$$

where the distribution function of background electrons, $f_{BG}(\vec{x},\vec{v},t)$, describes the evolution of the electrons' phase space density in the absence of the ion, and satisfies

$$\frac{\partial}{\partial t} f_{BG}(\vec{x},\vec{v},t) + \vec{v} \cdot \frac{\partial}{\partial \vec{x}} f_{BG}(\vec{x},\vec{v},t) + \vec{a} \cdot \frac{\partial}{\partial \vec{v}} f_{BG}(\vec{x},\vec{v},t) = 0, \quad (5)$$

with

$$\vec{a} \equiv -\frac{e}{m_e} \vec{E}_{ext} . \quad (6)$$

Since the acceleration does not depend either on the coordinate or on the initial velocity, the evolution of the distribution simply is a shift of the initial distribution by $\Delta \vec{v} = \vec{a}t$:

$$f_{BG}(\vec{x},\vec{v},t) = f_0(\vec{v} - \vec{a}t). \quad (7)$$

The solution in eq. (7) explicitly satisfies eq. (5). Inserting eq. (4) into eq. (2), and using eq. (5), lead to the linearized Vlasov equation

$$\frac{\partial}{\partial t} f_{ind}(\vec{x},\vec{v},t) + \vec{v} \cdot \frac{\partial}{\partial \vec{x}} f_{ind}(\vec{x},\vec{v},t) + \vec{a} \cdot \frac{\partial}{\partial \vec{v}} f_{ind}(\vec{x},\vec{v},t) + \frac{e}{m_e} \vec{\nabla} \phi_{ind}(\vec{x},t) \cdot \frac{\partial}{\partial \vec{v}} f_0(\vec{v} - \vec{a}t) = 0. \quad (8)$$

As the distribution of background electrons is uniform, and hence, does not contribute to the electric field, the electric potential, $\phi_{ind}(\vec{x},t)$, is determined solely by the modulation in electron density:

$$\nabla^2 \phi_{ind}(\vec{x},t) = -\frac{1}{\varepsilon_0} \left[ Z_i e \delta(\vec{x}) - e \int_{-\infty}^{\infty} f_{ind}(\vec{x},\vec{v},t) d^3v \right]. \quad (9)$$

Eqs. (8) and (9) constitute the linearized Vlasov-Poisson system that determines the electron phase space density modulation induced by the ion.

### III. ANALYTICAL SOLUTION FOR κ-2 VELOCITY DISTRIBUTION

To proceed, it is convenient to change the independent variables, $\vec{v}$ and $t$, to a set of new ones:

$$\vec{u} \equiv \vec{v} - \vec{a}t , \quad (10)$$

and,

$$\tau \equiv t . \quad (11)$$

We denote the induced variation in the phase space density in terms of the new variables as follows:

$$f_1(\vec{x},\vec{u},\tau) \equiv f_{ind}(\vec{x},\vec{u}+\vec{a}\tau,\tau). \tag{12}$$

With the new variables, the partial derivatives in eq. (8) with respect to $\vec{v}$ and $t$ can be rewritten as

$$\left.\frac{\partial}{\partial t} f_{ind}(\vec{x},\vec{v},t)\right|_{\vec{v}} = -\left.\frac{\partial}{\partial \vec{u}} f_1(\vec{x},\vec{u},\tau)\right|_{\tau} \cdot \vec{a} + \left.\frac{\partial}{\partial \tau} f_1(\vec{x},\vec{u},\tau)\right|_{\vec{u}}, \tag{13}$$

and,

$$\left.\vec{a}\cdot\frac{\partial}{\partial \vec{v}} f_{ind}(\vec{x},\vec{v},t)\right|_{t} = \vec{a}\cdot\left.\frac{\partial}{\partial \vec{u}} f_1(\vec{x},\vec{u},\tau)\right|_{\tau}. \tag{14}$$

Inserting eqs. (13) and (14) into eq. (8) yields the linearized Vlasov equation in the new variables:

$$\frac{\partial}{\partial \tau} f_1(\vec{x},\vec{u},\tau) + (\vec{u}+\vec{a}\tau)\cdot\frac{\partial}{\partial \vec{x}} f_1(\vec{x},\vec{u},\tau) + \frac{e}{m_e}\vec{\nabla}\phi_{ind}(\vec{x},\tau)\cdot\frac{\partial}{\partial \vec{u}} f_0(\vec{u}) = 0. \tag{15}$$

The Poisson equation, eq. (9), can be rewritten simply as

$$\nabla^2 \phi_{ind}(\vec{x},\tau) = -\frac{1}{\varepsilon_0}\left[Z_i e\delta(\vec{x}) - e\int_{-\infty}^{\infty} f_1(\vec{x},\vec{u},\tau)d^3u\right]. \tag{16}$$

Multiplying both sides of eqs. (15) and (16) by $e^{-i\vec{k}\cdot\vec{x}}$, and integrating over $\vec{x}$ gives their Fourier transformation:

$$\frac{\partial}{\partial \tau}\tilde{f}_1(\vec{k},\vec{u},\tau) + i\vec{k}\cdot(\vec{u}+\vec{a}\tau)\tilde{f}_1(\vec{k},\vec{u},\tau) = -i\frac{e}{m_e}\tilde{\phi}_{ind}(\vec{k},\tau)\vec{k}\cdot\frac{\partial}{\partial \vec{u}} f_0(\vec{u}), \tag{17}$$

and,

$$\tilde{\phi}_{ind}(\vec{k},\tau) = \frac{e}{\varepsilon_0 k^2}\left[Z_i - \tilde{n}_1(\vec{k},\tau)\right]. \tag{18}$$

The Fourier components of the phase space density and electric potential are defined as

$$\tilde{f}_1(\vec{k},\vec{u},\tau) = \int_{-\infty}^{\infty} f_1(\vec{x},\vec{u},\tau)e^{-i\vec{k}\cdot\vec{x}}d^3x, \tag{19}$$

$$\tilde{\phi}_{ind}(\vec{k},\tau) = \int_{-\infty}^{\infty} \phi_{ind}(\vec{x},\tau)e^{-i\vec{k}\cdot\vec{x}}d^3x, \tag{20}$$

and the Fourier components of the modulation in spatial density is given by

$$\tilde{n}_1(\vec{k},\tau) = \int_{-\infty}^{\infty} \tilde{f}_1(\vec{k},\vec{u},\tau)d^3u. \tag{21}$$

Multiplying both sides of eq. (17) by $\exp\left(i\vec{k}\cdot\vec{u}\tau + i\vec{k}\cdot\vec{a}\frac{\tau^2}{2}\right)$, it becomes

$$\frac{\partial}{\partial \tau}\left[e^{i\vec{k}\cdot\left(\vec{u}+\frac{\vec{a}\tau}{2}\right)\tau}\tilde{f}_1\left(\vec{k},\vec{u},\tau\right)\right] = -i\frac{e}{m_e}\left[\vec{k}\cdot\frac{\partial}{\partial \vec{u}}f_0\left(\vec{u}\right)\right]e^{i\vec{k}\cdot\left(\vec{u}+\frac{\vec{a}\tau}{2}\right)\tau}\tilde{\phi}_{ind}\left(\vec{k},\tau\right), \quad (22)$$

that, after integration over $\tau$, produces

$$\tilde{f}_1\left(\vec{k},\vec{u},\tau\right) = -i\frac{e}{m_e}\int_0^\tau \tilde{\phi}_{ind}\left(\vec{k},\tau_1\right)e^{\frac{i\vec{k}\cdot\vec{a}\left(\tau_1^2-\tau^2\right)}{2}}\left[\vec{k}\cdot\frac{\partial}{\partial \vec{u}}f_0\left(\vec{u}\right)\right]e^{i\vec{k}\cdot\vec{u}(\tau_1-\tau)}d\tau_1. \quad (23)$$

Inserting eq. (18) into (23), and then taking the integration over $\vec{u}$, leads to the following integral equation

$$\tilde{n}_1\left(\vec{k},\tau\right) = \frac{e^2}{\varepsilon_0 m_e}\int_0^\tau d\tau_1 \left[\tilde{n}_1\left(\vec{k},\tau_1\right)-Z_i\right]e^{\frac{i\vec{k}\cdot\vec{a}\left(\tau_1^2-\tau^2\right)}{2}}\int_{-\infty}^\infty \left[\frac{i\vec{k}}{k^2}\cdot\frac{\partial}{\partial \vec{u}}f_0\left(\vec{u}\right)\right]e^{i\vec{k}\cdot\vec{u}(\tau_1-\tau)}d^3u. \quad (24)$$

To move further, we assume that the initial distribution of the velocity of electrons at $t=0$ is an anisotropic κ-2 distribution, which, in terms of the new variables, gives the distribution of background electrons as follows:

$$f_0\left(\vec{u}\right) = \frac{n_0}{\pi^2 \beta_x \beta_y \beta_z}\left[1+\frac{\left(u_x+v_{0x}\right)^2}{\beta_x^2}+\frac{\left(u_y+v_{0y}\right)^2}{\beta_y^2}+\frac{\left(u_z+v_{0z}\right)^2}{\beta_z^2}\right]^{-2}, \quad (25)$$

where $\vec{v}_0$ is the velocity of the ion, and $n_0$ is the spatial density of the background electrons. Parameters $\beta_x$, $\beta_y$ and $\beta_z$ describe the velocity spreads of electrons in the corresponding directions. Inserting eq. (25) into (24) and applying the relation

$$i\int_{-\infty}^\infty e^{i\vec{k}\cdot\vec{u}\tau}\frac{\vec{k}}{k^2}\cdot\frac{\partial}{\partial \vec{u}}f_0\left(\vec{u}\right)d^3u = \int_{-\infty}^\infty f_0\left(\vec{u}\right)e^{i\vec{k}\cdot\vec{u}\tau}\tau d^3u, \quad (26)$$

the integral equation is reduced to

$$\begin{aligned}\tilde{n}_1\left(\vec{k},\tau\right) &= \omega_p^2 \int_0^\tau \left[\tilde{n}_1\left(\vec{k},\tau_1\right)-Z_i\right]e^{\frac{i\vec{k}\cdot\vec{a}\left(\tau_1^2-\tau^2\right)}{2}}\left(\tau_1-\tau\right)g\left(\vec{k}\left(\tau-\tau_1\right)\right)d\tau_1 \\ &= \omega_p^2 \int_0^\tau \left[\tilde{n}_1\left(\vec{k},\tau_1\right)-Z_i\right]e^{\frac{i\vec{k}\cdot\vec{a}\left(\tau_1^2-\tau^2\right)}{2}}\left(\tau_1-\tau\right)e^{\lambda(\vec{k})(\tau-\tau_1)}d\tau_1\end{aligned}, \quad (27)$$

where

$$g\left(\vec{q}\right) \equiv \frac{1}{n_0}\int_{-\infty}^\infty f_0\left(\vec{u}\right)e^{-i\vec{q}\cdot\vec{u}}d^3u = \exp\left[i\vec{q}\cdot\vec{v}_0-\sqrt{q_x^2\beta_x^2+q_y^2\beta_y^2+q_z^2\beta_z^2}\right], \quad (28)$$

$$\lambda\left(\vec{k}\right) \equiv i\vec{k}\cdot\vec{v}_0-\sqrt{k_x^2\beta_x^2+k_y^2\beta_y^2+k_z^2\beta_z^2}, \quad (29)$$

and,

$$\omega_p \equiv \sqrt{\frac{n_0 e^2}{m_e \varepsilon_0}} \ . \tag{30}$$

Eq. (27) can be written into a more compact form:

$$\tilde{H}_1(\vec{k},\tau) = \omega_p^2 \int_0^\tau \left[ \tilde{H}_1(\vec{k},\tau_1) - Z_i e^{\frac{i\vec{k}\cdot\vec{a}\tau_1^2}{2} - \lambda(\vec{k})\tau_1} \right] (\tau_1 - \tau) d\tau_1 \ , \tag{31}$$

and, the new function, $\tilde{H}_1(\vec{k},\tau)$, is defined as

$$\tilde{H}_1(\vec{k},\tau) \equiv \tilde{n}_1(\vec{k},\tau) e^{-\lambda(\vec{k})\tau + \frac{i\vec{k}\cdot\vec{a}\tau^2}{2}} \ . \tag{32}$$

Taking the second-time derivative of eq. (31) generates an inhomogeneous second-order ordinary differential equation (ODE)

$$\frac{d^2}{d\tau^2} \tilde{H}_1(\vec{k},\tau) + \omega_p^2 \tilde{H}_1(\vec{k},\tau) = Z_i \omega_p^2 e^{\frac{i\vec{k}\cdot\vec{a}\tau^2}{2} - \lambda(\vec{k})\tau} \ , \tag{33}$$

which, for arbitrary initial conditions, has the solution [6]

$$\tilde{H}_1(\vec{k},\tau) = c_1 \cos(\omega_p \tau) + c_2 \sin(\omega_p \tau) + Z_i \omega_p \int_0^\tau \sin\left[\omega_p(\tau - \tau_1)\right] e^{\frac{i\vec{k}\cdot\vec{a}\tau_1^2}{2} - \lambda(\vec{k})\tau_1} d\tau_1 \ , \tag{34}$$

with $c_1$ and $c_2$ being constants to be determined by the initial conditions at $t = 0$. As we assume that there is no modulation at $t = 0$, the initial conditions read

$$\tilde{H}_1(\vec{k},0) = \tilde{n}_1(\vec{k},0) = 0 \ , \tag{35}$$

and,

$$\left. \frac{d}{d\tau} \tilde{H}_1(\vec{k},\tau) \right|_{\tau=0} = \left. \frac{d}{d\tau} \tilde{n}_1(\vec{k},\tau) \right|_{\tau=0} - \lambda(\vec{k}) \tilde{n}_1(\vec{k},0) = 0. \tag{36}$$

Applying the initial conditions of eq. (35) to eq. (34) for $\tau = 0$ yields

$$c_1 = 0 \ . \tag{37}$$

Inserting eq. (37) into (34) and then taking the first-time derivative produces

$$\frac{d}{d\tau} \tilde{H}_1(\vec{k},\tau) = c_2 \omega_p \cos(\omega_p \tau) + Z_i \omega_p^2 \cos(\omega_p \tau) \int_0^\tau \cos(\omega_p \tau_1) e^{\frac{i\vec{k}\cdot\vec{a}\tau_1^2}{2} - \lambda(\vec{k})\tau_1} d\tau_1$$
$$+ Z_i \omega_p^2 \sin(\omega_p \tau) \int_0^\tau \sin(\omega_p \tau_1) e^{\frac{i\vec{k}\cdot\vec{a}\tau_1^2}{2} - \lambda(\vec{k})\tau_1} d\tau_1 \quad . \tag{38}$$

Eqs. (36) and (38) require

$$c_2 = 0, \tag{39}$$

and hence, we obtain from eqs. (34), (37), and (39)

$$\tilde{H}_1(\vec{k},\tau) = Z_i \omega_p \int_0^\tau \sin\left[\omega_p(\tau-\tau_1)\right] e^{-\lambda(\vec{k})\tau_1 + \frac{i\vec{k}\cdot\vec{a}\tau_1^2}{2}} d\tau_1 \ . \tag{40}$$

Substituting the definition of $\tilde{H}_1(\vec{k},\tau)$, eq. (32), back into eq. (40), we obtain the modulation in electron density in the wave-vector domain

$$\tilde{n}_1(\vec{k},\tau) = Z_i \omega_p \int_0^\tau \sin\left[\omega_p(\tau-\tau_1)\right] e^{\lambda(\vec{k})(\tau-\tau_1) - \frac{i\vec{k}\cdot\vec{a}(\tau^2-\tau_1^2)}{2}} d\tau_1. \tag{41}$$

The electron-density modulation in the configuration space is given by the inverse Fourier transformation of $\tilde{n}_1(\vec{k},\tau)$, i.e.,

$$n_1(\vec{x},t) = \frac{1}{(2\pi)^3} \int_{-\infty}^\infty \tilde{n}_1(\vec{k},t) e^{i\vec{k}\cdot\vec{x}} d^3k \ . \tag{42}$$

Inserting eq. (41) into eq. (42), we finally obtain the following expression for the electron-density modulation induced by an ion

$$\begin{aligned}n_1(\vec{x},t) = \frac{Z_i}{\pi^2 r_x r_y r_z} \int_0^{\omega_p t} \psi \sin\psi \Bigg[ \psi^2 &+ \left(\bar{x} - \bar{a}_x \psi\left(\omega_p t - \frac{\psi}{2}\right) + \bar{v}_{0,x}\psi\right)^2 \\ &+ \left(\bar{y} - \bar{a}_y \psi\left(\omega_p t - \frac{\psi}{2}\right) + \bar{v}_{0,y}\psi\right)^2 + \left(\bar{z} - \bar{a}_z \psi\left(\omega_p t - \frac{\psi}{2}\right) + \bar{v}_{0,z}\psi\right)^2 \Bigg]^{-2} d\psi\end{aligned} \tag{43}$$

wherein we used the normalized variables, defined as $\bar{x}_j \equiv x_j/r_j$, $\bar{a}_j \equiv a_j/r_j\omega_p^2$, $\bar{v}_{0,j} = v_{0,j}/\beta_j$ and $r_j \equiv \beta_j/\omega_p$ for $j = x, y, z$. Eq. (43) has the form of a 1-D integral with finite integration range, and, as expected, it reduces to the previously derived results at the limit of $\vec{a} = 0$ [5]. Figs. 1 and 2 show the 1-D and 2-D plots of the modulation of electron density obtained by the numerical evaluation of eq. (43). In Fig. 1, we plot the modulations in electron density at a specific transverse location for various longitudinal space charge fields. For an ion at rest, as seen in Figs. 1(a) and 2(a), the space charge field reduces the amplitude of the peak modulation and shifts its longitudinal location. However, for a moving ion, Figs. 1(b) and 2(b) show that the acceleration of electrons due to space charge field can compensate for the effects due to ion motion provided that the space charge force is in the same direction as the ion's velocity. Qualitatively, this can be understood as a matching between the hadrons' velocity and average velocity of the electrons during their interaction. Hence, matching average electron's velocity with that of the ion should increase the amplitude of modulation. Direct numerical evaluation of eq. (43) showed that the effect nearly is compensated (within a deviation of a few percent) when the normalized velocity and the acceleration is matched. Fig. 3 illustrates such compensation for three phase-advances of the plasma oscillations. The matching naturally depends on the phase advance. At phase advance $\omega_p t = \pi/4$, the matching occurs at about $\bar{v}_z \approx 0.63 \bar{a}_z$. For phase advances of $\omega_p t = \pi/2$ and $\omega_p t = \pi$ the matching ratios are $\bar{v}_z \approx 1.35 \bar{a}_z$ and $\bar{v}_z \approx 2.9 \bar{a}_z$, correspondingly.

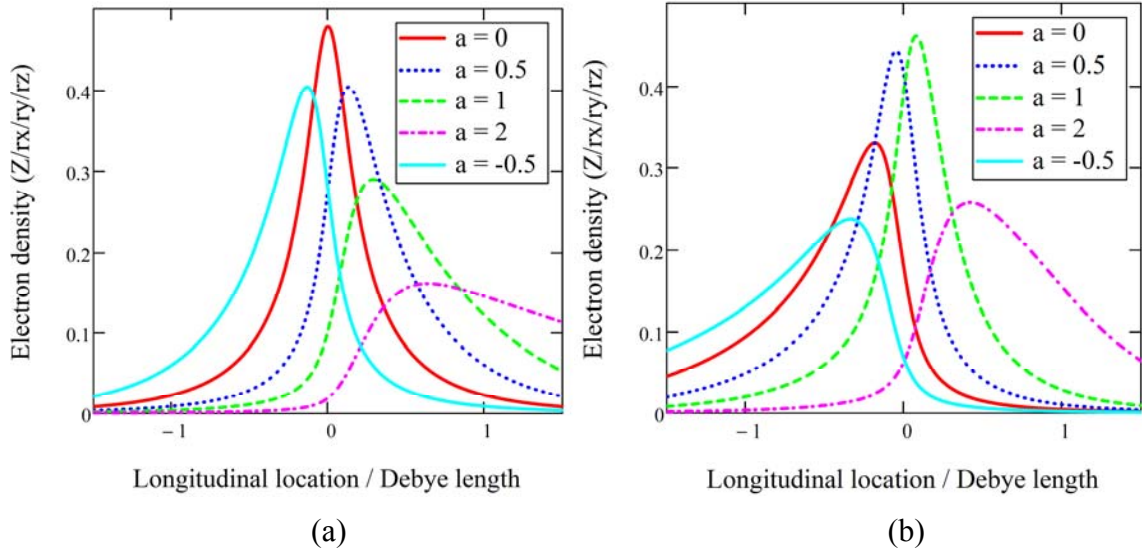

Figure 1. Profiles of the density modulation induced by an ion in the presence of an external electric field (as calculated in eq. (43)). The external electric field is along the z direction and the following values are used for the normalized acceleration, $\bar{a}_z$: 0 (red), 0.5 (blue), 1 (green), 2 (magenta), and -0.5 (light blue). The abscissa is the longitudinal location in units of longitudinal Debye length, $r_z$, and the ordinate is the electron density at transverse location $x = 0.1 r_x$ and $y = 0.1 r_y$ in units of $Z_i / (r_x r_y r_z)$. The snapshot is taken at $\omega_p t = \pi/2$. (a) The ion is at rest; (b) the ion is moving with velocity $v_{0,z} = \beta_z$.

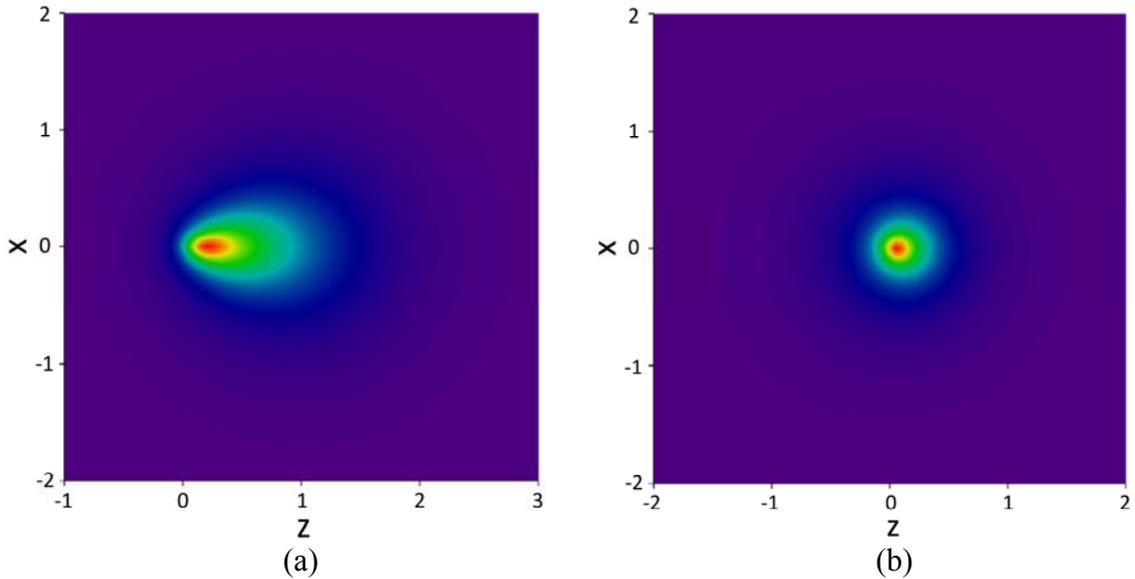

Figure 2. Modulation in electron density induced by an ion in the presence of an external electric field in the z direction with $\bar{a}_z = 1$. The abscissa is the location along the z direction in units of the longitudinal Debye length, $r_z$, and the ordinate is the location along x in units of the Horizontal Debye length, $r_x$. (a) The ion is at rest; (b) the ion is moving at $v_{0,z} = \beta_z$. The snapshot is taken at $\omega_p t = \pi/2$ and $y = 0.1 r_y$.

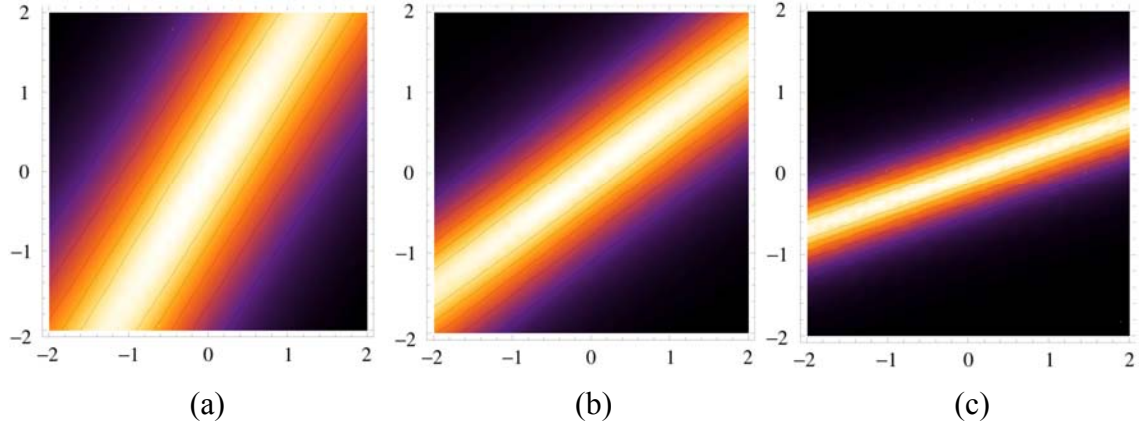

(a)                          (b)                          (c)

Figure 3. Plots of the normalized density at $z = 0$, $x = 0.1 r_x$ and $y = 0.1 r_y$ as functions of $\bar{v}_z$ (horizontal axis), and $\bar{a}_z$ (vertical axis) for three-phase advances of plasma oscillations: (a) $\omega_p t = \pi/4$; (b) $\omega_p t = \pi/2$; and, (c) $\omega_p t = \pi$. The distributions are normalized to their maximum values and the contour lines are spaced by 0.2.

As FEL only amplifies modulation in electron current with frequencies close to its resonant frequency, the following quantity is closely related to the efficiency of modulation :

$$\eta(k_z, t) \equiv \int_{-\infty}^{\infty} dz\, e^{-ik_z z} \int_{-\infty}^{\infty} n_1(x, y, z, t) dx dy . \tag{44}$$

Inserting eq. (42) into (44) leads to

$$\eta(k_z, t) = \frac{1}{2\pi} \int_{-\infty}^{\infty} dk_{1,z} \tilde{n}_1(0, 0, k_{1,z}, t) \int_{-\infty}^{\infty} e^{i(k_{1,z} - k_z)z} dz = \tilde{n}_1(0, 0, k_z, t) \tag{45}$$

Using eq. (41), then eq. (45) becomes

$$\eta(k_z, t) = Z_i \int_0^{\omega_p t} \exp\left[-i\bar{k}_z \bar{a}_z \psi\left(\omega_p t - \frac{\psi}{2}\right) + \left(i\bar{k}_z \cdot \bar{v}_{0,z} - |\bar{k}_z|\right)\psi\right] \sin\psi\, d\psi . \tag{46}$$

Fig. 4 plots the amplitude and phase of $\eta(k_z, t)$ for various acceleration parameters as a function of $k_z r_z$. As shown in Fig. 4(a), for $|a_z| \leq r_z \omega_p^2$ and the FEL resonant wavelength $\lambda_{FEL} \geq 2\pi r_z$, the changes in amplitude due to the longitudinal space charge is negligible. However, Fig. 4(b) shows that a considerable phase shift can occur, even for a modest acceleration due to the space charge field. To qualitatively understand the impact of longitudinal space charge field on the CeC modulation process, as examples, we continue with numerical calculations for a few proposed CeC schemes.

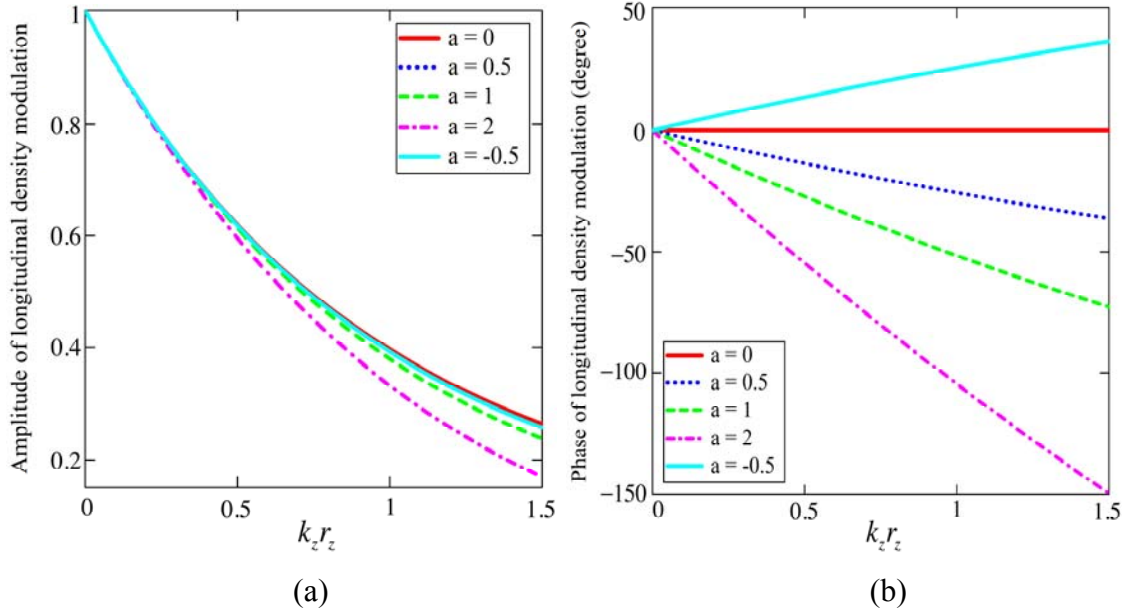

(a)                                (b)

Figure 4: Fourier components of the modulation in longitudinal density calculated from eq. (46) for various normalized acceleration parameters. The abscissa is the normalized longitudinal wave number, $k_z r_z$, and the plots are taken at $\omega_p t = \pi/2$. (a) the ordinate is $|\eta(k_z,t)|/Z_i$; (b) the ordinate is the phase shift of $\eta(k_z,t)$ with respect to that of the zero-acceleration parameter.

## IV. NUMERICAL EXAMPLES

The CeC proof of principle (CeC PoP) experiment is under construction at BNL [7] and a few possible CeC designs have been proposed [8]. The designed parameters of the electron beam for three CeCs are listed in Table 1 [7, 8]. In this section, we estimate the effects of the longitudinal long-range space charge field on the modulation process for these parameters.

Table 1: Electron beam parameters for the Proof of Principle Experiment of CeC

| Parameter/CeC | CeC PoP | eRHIC CEC | LHC CeC |
|---|---|---|---|
| Bunch charge, nC | 1 | 10 | 30 |
| Bunch length, rms, beam frame, $\sigma_z$, m | 0.126 | 6.3 | 893 |
| Beam radius R, mm | 1.3 | 0.35 | 0.15 |
| R/$\sigma_z$ | $1.03 \cdot 10^{-2}$ | $5.56 \cdot 10^{-5}$ | $1.68 \cdot 10^{-7}$ |
| Energy (MeV)/ γ | 21.5 / 42 | 136.2 / 266 | 3,812 / 7,460 |
| Longitudinal Debye length at the bunch center, beam frame, μm | 42 | 19.1 | 21.7 |
| Plasma phase advance in modulator, rad | $\pi/2$ | $\pi/2$ | 0.062 |

Space charge effects are known to fall very fast with the energy of the particles. Hence, we first consider the effects of space charge for the CeC PoP experiment and latter make estimates for two other cases.

First, we calculate the longitudinal space charge field inside the electron bunch. For simplicity, we only calculate the field at the bunch axis, i.e., for $x = y = 0$. In addition, we assume the electron bunch has beer-can transverse distribution, i.e., the electron density is uniform for $r \leq R$ and zero for $r > R$. The system has cylindrical symmetry, and hence, it is more convenient to use cylindrical coordinates. As illustrated in Fig. 5, the longitudinal space charge field at location $(0,0,l)$ contributed by electrons in an infinitesimal volume at location $(r,\varphi,\zeta)$ is given by[†]

$$\Delta E_z(l;r,\varphi,\zeta) = \frac{r(l-\zeta)\rho(\zeta)\theta(R-r)\Delta\zeta\Delta r\Delta\varphi}{4\pi\varepsilon_0\left[(\zeta-l)^2+r^2\right]^{\frac{3}{2}}}, \qquad (47)$$

where $\theta(x)$ is the Heaviside step function with the definition

$$\theta(x) \equiv \begin{cases} 1, & x \geq 0 \\ 0, & x < 0 \end{cases}, \qquad (48)$$

and $\rho(\zeta)$ is the density of the electron charge in the beam frame for $r \leq R$. Integrating eq. (47) over the transverse area of the beam yields the space charge field at location $(0,0,l)$ due to a longitudinal slice of electrons at longitudinal location $\zeta$ with width $\Delta\zeta$

$$\Delta E_z(l;\zeta) = \frac{\rho(\zeta)\Delta\zeta}{2\varepsilon_0}\left[\frac{l-\zeta}{|l-\zeta|} - \frac{l-\zeta}{\sqrt{(l-\zeta)^2+R^2}}\right]. \qquad (49)$$

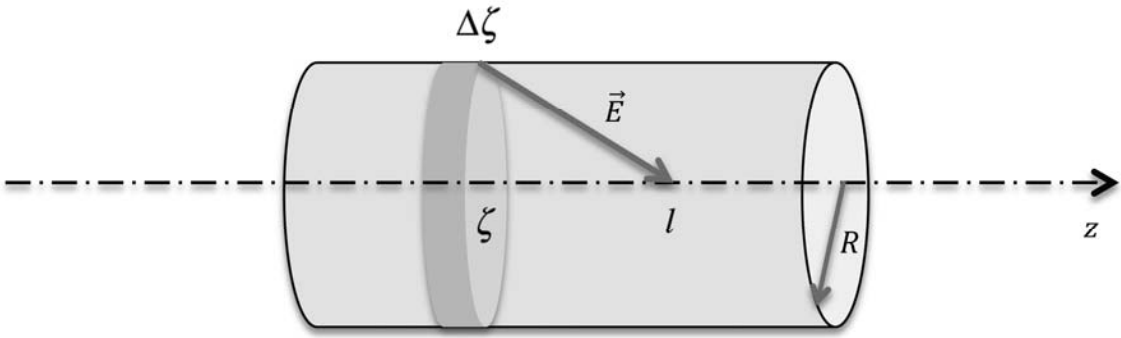

Figure 5. Illustration of our calculation of the longitudinal space charge field.

---

[†] In this section, we use the notation, $l$ to represent the global longitudinal location along the bunch where space charge varies substantially, while leaving $z$ for the local longitudinal coordinate where the characteristic scale is the longitudinal Debye length, and the variation of space charge field is negligible.

To proceed, we assumed that the electrons' charge distribution at $r \leq R$ has the following form

$$\rho(\zeta) = \frac{Q_e}{\pi R^2} \frac{1}{\sqrt{2\pi}\sigma_z} e^{-\frac{\zeta^2}{2\sigma_z^2}}, \quad (50)$$

where $Q_e$ is the total charge of the electron bunch, and $\sigma_z$ is the R.M.S. length of the electron bunch. Inserting eq. (50) into (49) and then integrating over $\zeta$ yields the longitudinal space charge field at the location $(0,0,l)$,

$$E_z(l) = \frac{-Q_e}{\sqrt{2\pi^{\frac{3}{2}}}\varepsilon_0 \sigma_z^2} \cdot F\left(\frac{l}{\sigma_z}, \frac{R}{\sigma_z}\right), \quad (51)$$

with,

$$F(\zeta, \chi) \equiv \frac{e^{-\frac{\zeta^2}{2}}}{\chi^2} \cdot \int_0^\infty e^{-\frac{\xi^2}{2}} \sinh(\zeta \cdot \xi)\left(\frac{\xi}{\sqrt{\xi^2 + \chi^2}} - 1\right) d\xi. \quad (52)$$

We note that, as shown in Fig. 6, even though the values of $R/\sigma_z$ varies by five orders-of-magnitude for the three cases listed in Table 1, the peak values of $F(l/\sigma_z, R/\sigma_z)$ only change by a factor of four. We numerically evaluated eq. (51) for the CeC PoP parameters and plot the results in Fig. 10 (blue); they show the maximal longitudinal space charge field reaches about 1.5 KV/m.

As detailed in the previous section, the effects of the space charge relate to the normalized acceleration parameter, $\bar{a}_z$, which can be calculated as follows:

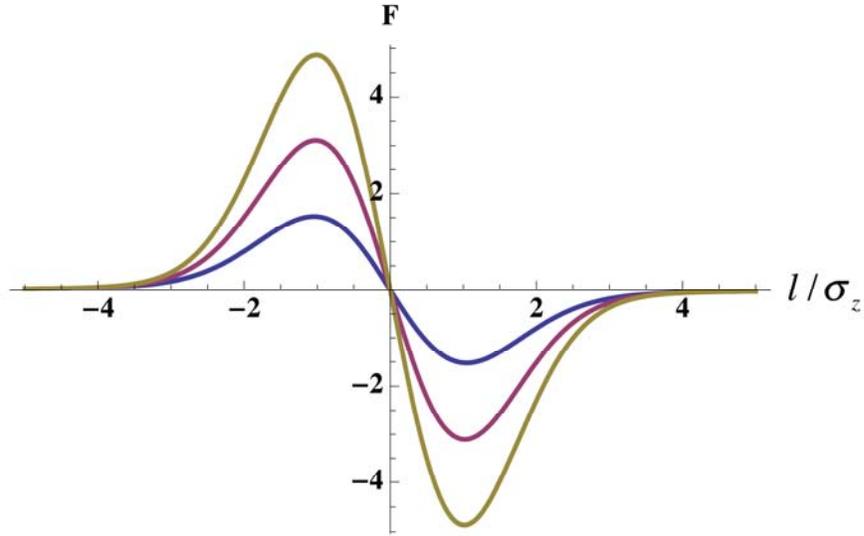

Figure 6. Plot of the F function for three $R/\sigma_z$ parameters in Table 1 versus $l/\sigma_z$. Blue is for the CeC PoP case, magenta is for eRHIC CeC, and grey-green is for LHC CeC.

$$\bar{a}_z(l) = \frac{-eE_z(l)}{m_e \beta_z \omega_p(l)} = \frac{-eE_z(l)}{m_e r_z(0) \omega^2_p(0)} \cdot e^{\frac{l^2}{4\sigma_z^2}}, \qquad (53)$$

where,

$$\omega_p(0) = \sqrt{\frac{|Q_e|e}{\sqrt{2\pi^3} R^2 \sigma_z m_e \varepsilon_0}} \qquad (54)$$

and,

$$r_z(0) = \frac{\beta_z}{\omega_p(0)} \qquad (55)$$

are the plasma frequency and the longitudinal Debye length at the center of the electron bunch. The plasma frequency at location $l$ is

$$\omega_p(l) = e^{-\frac{l^2}{4\sigma_z^2}} \omega_p(0) \qquad (56)$$

Employing eqs. (51) and (53), we obtain the expression for the normalized acceleration parameter as follows:

$$\bar{a}_z(l) = \frac{R^2}{r_z(0) \cdot \sigma_z} e^{\frac{l^2}{4\sigma_z^2}} \cdot F\left(\frac{l}{\sigma_z}, \frac{R}{\sigma_z}\right). \qquad (57)$$

Fig. 7 plots the normalized acceleration parameter along the electron bunch for the CeC PoP parameters, suggesting that it stays below one within $\pm 4\sigma_z$ of the bunch.

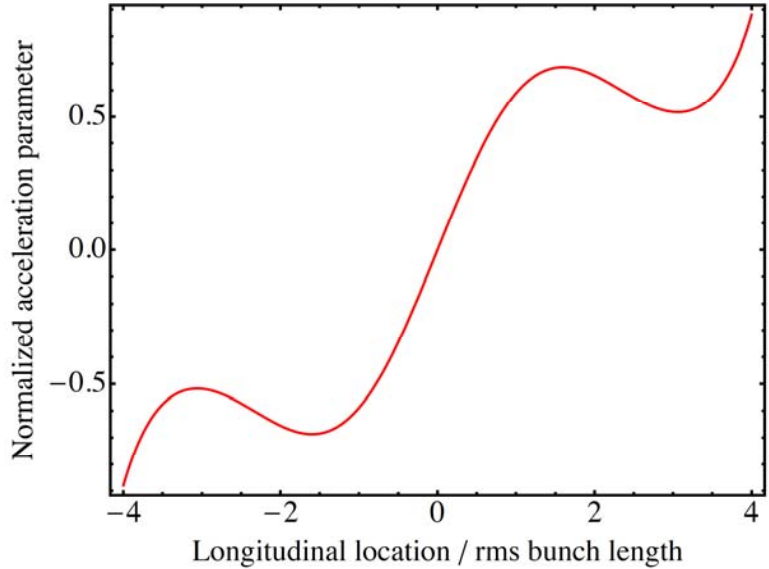

Figure 7: Normalized acceleration parameter due to space charge field (for the CeC PoP parameters listed in Table 1) as function of the longitudinal coordinate within a bunch.

The apparent growth of $|\bar{a}_z(l)|$ at $|l|/\sigma_z > 3$ in Fig. 7 does not mean that the effects of space charge field on the modulation are stronger for electrons far away from the bunch's center. In practice, the time of interaction $t = \varphi_p(0)/\omega_p(0)$ is fixed, and not the local phase advances of plasma oscillations $\varphi_p(l) = \omega_p(l)t$, with $l$ being the distance from the bunch's center. Hence, to properly evaluate the space charge effect at $l \neq 0$, we must evaluate eq. (43) at

$$\omega_p(l)t = \varphi_p(0)\exp(-l^2/4\sigma_l^2). \tag{58}$$

Fig. 8 illustrates the dependence of the density modulation on $\bar{a}_z$ as function of the phase of plasma oscillation, e.g., eq. (58). It shows that at small phase advances, the dependence is very weak. It is apparent from Fig. 7 that $\bar{a}_z$ reaches a local extreme of $|\bar{a}_z| \sim 0.7$ at $|z|/\sigma_z \approx 1.75$. At this location, the plasma frequency is two-times smaller than that in the beam's center, and with $\varphi_p(0) = \pi/2 \to \varphi_p(1.75\sigma_z) \approx \pi/4$. Hence, the effect on the peak density does not exceed 10 percent for $|\bar{a}_z| \sim 0.7$. In addition, at $|z|/\sigma_z \approx 1.75$, the e-beam's peak current is about 1/5$^{th}$ of that in the center. Therefore, in the FEL-based CeC, the FEL gain is turned off at this location, and hence, this part of the beam does not effectively participate in the cooling process.

To account for the variation of $\omega_p(l)$ along the bunch, whilst estimating the influences of the longitudinal space's charge field on the efficiency of modulation, we rewrite eq. (46) in non-normalized variables:

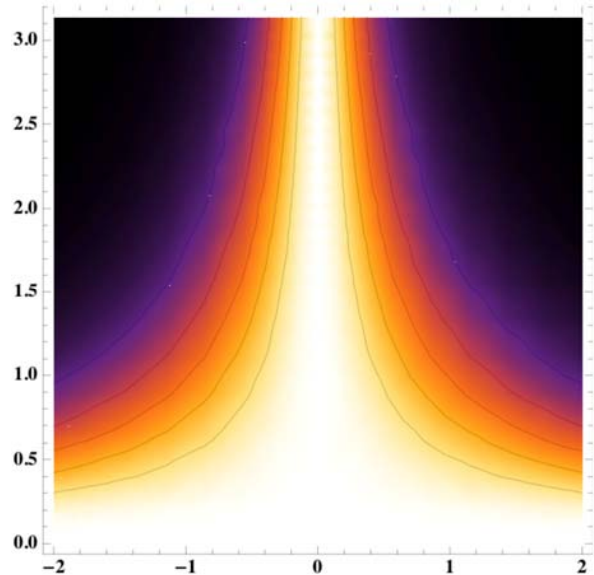

Figure 8. Plots of the normalized density at $z = 0$, $x = 0.1r_x$ and $y = 0.1r_y$ as functions of $\bar{a}_z$ (horizontal axis), and $\varphi_p = \omega_p t$ (vertical axis) $\omega_p t = \pi$. The distributions are normalized to their values at $\bar{a}_z = 0$. The contour lines are spaced by 0.2.

$$\eta(k_z,t,l) = Z_i\omega_p(l)\int_0^t \exp\left[-ik_z a_z(l)\left(t-\frac{\tau}{2}\right)\tau + \left(ik_z \cdot v_{0,z} - |k_z|\beta_z\right)\tau\right]\sin(\omega_p(l)t)d\tau, \tag{59}$$

and the space charge influences along the electron bunch can be represented by the following quantities:

$$\Delta\eta_{amp}(k_z,t,l) \equiv \frac{|\eta(k_z,t,l)| - |\eta_0(k_z,t,l)|}{|\eta_0(k_z,t,l)|} \tag{60}$$

and,

$$\Delta\eta_{ph}(k_z,t,l) \equiv \arg[\eta(k_z,t,l)] - \arg[\eta_0(k_z,t,l)] \tag{61}$$

where,

$$\eta_0(k_z,t,l) = Z_i\omega_p(l)\int_0^t \exp\left[\left(ik_z \cdot v_{0,z} - |k_z|\beta_z\right)\tau\right]\sin(\omega_p(l)t)d\tau \tag{62}$$

is the Fourier components in the absence of the space charge effects, i.e., $a_z = 0$. Fig. 9 plots the relative changes of $\eta(k_z,t,l)$ calculated from eqs. (60) and (61) for the experimental proof of CeC- principle .

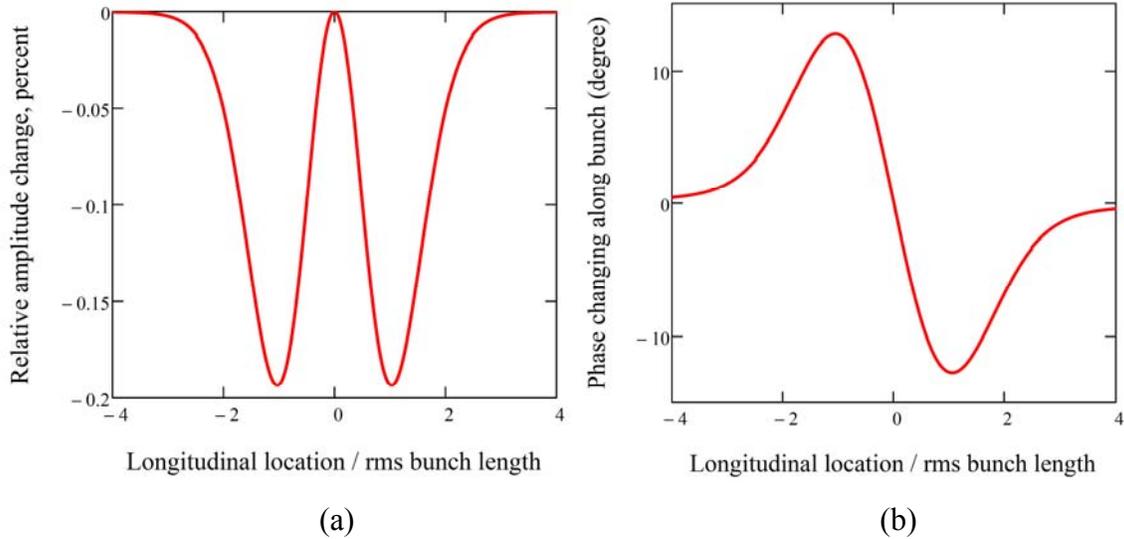

(a)          (b)

Figure 9: Influences of the longitudinal space charge field on the Fourier components of the longitudinal density-modulation at the FEL resonant wavelength as a function of longitudinal location along the electron bunch. The ion is at rest, and the parameters from the proof of the CeC principle experiment are applied in generating these plots. The abscissa is the longitudinal location in units of R.M.S. bunch length. (a) the relative change in percentage of the amplitude of the Fourier component calculated from eq. (60); and, (b) the phase change of Fourier component in degrees, calculated from eq. (61).

As shown in Fig. 9(a), the reduction in amplitude of the wave-packets due to the longitudinal space charge effects at the modulator is under 0.2%. However, Fig. 9(b) shows that the phase change of the initial modulation will result in the maximal wave-packet phase-shift of ±13 degrees. Since the reduction of the CeC efficiency is proportional to the cosine of phase shift, this effect would lower CeC efficiency by less than 3%.

Compared with the CeC proof-of-principle experiment, the longitudinal space charge fields for the other two cases listed in the Table 1 are dramatically lower. In the eRHIC CeC scheme, the strength of the space charge field peaks at 6.15 V/m, while for the LHC CeC scheme, it is about 0.001 V/m. Consequently, the peak value of $\bar{a}_z$ falls, to $\pm\,0.003$ and $\pm\,3.7\times10^{-6}$, respectively, for the proposed eRHIC- and LHC-CeC schemes.

## V. SCREENING EFFECTS FROM THE BEAM PIPE

In reality, the electron bunch usually is enclosed by metallic vacuum chamber, the walls of which can reduce the strength of the longitudinal field induced by the space charge. For $\sigma_z/b \gg 1$, the longitudinal space charge field in terms of the beam frame variables is given by[‡] [9]

$$E_{scr,z}(z) = \frac{e}{4\pi\varepsilon_0}\left[2\ln\left(\frac{b}{R}\right)+1\right]\frac{d\lambda}{dz}, \tag{63}$$

where $b$ is the beam pipe's radius, and

$$\lambda(z) = -\frac{\pi R^2}{e}\rho(z) \tag{64}$$

is the electrons' line number density. Inserting eqs. (50) and (64) into eq. (63) yields

$$E_{app,z}(z) = \frac{Q_e}{4\sqrt{2}\pi^{\frac{3}{2}}\varepsilon_0\sigma_z^2}\left[2\ln\left(\frac{b}{R}\right)+1\right]\frac{z}{\sigma_z}e^{-\frac{z^2}{2\sigma_z^2}}. \tag{65}$$

More generally, in the presence of a circular perfectly-conducting beam-pipe, the on-axis longitudinal space charge field of an electron bunch with the distribution of eq. (50) and arbitrary bunch length is given by the following 1-D integral (Appendix A):

$$E_{exa,z}(0,z) = -\frac{Q_e}{\varepsilon_0\pi^2 R}\int_0^\infty e^{-\frac{k_z^2\sigma_z^2}{2}}\left[\frac{I_1(k_zR)K_0(k_zb)}{I_0(k_zb)}+K_1(k_zR)-\frac{1}{k_zR}\right]\sin(k_zz)dk_z, \tag{66}$$

where $I_n(x)$ and $K_n(x)$ are the modified Bessel functions. In Fig 10, we plot the space charge field calculated from eqs. (65) and (66) for the parameters of the proof of CeC principle experiment; they show that the formulae agree well for the considered parameters. More importantly, this figure implies that the shielding effects from a perfectly conducting beam pipe wall reduce by 20% the peak field from the longitudinal space charge.

---
[‡] We defined $e \equiv 1.602...\times 10^{-19}\,C$ across this article, which is different from the definition in [9].

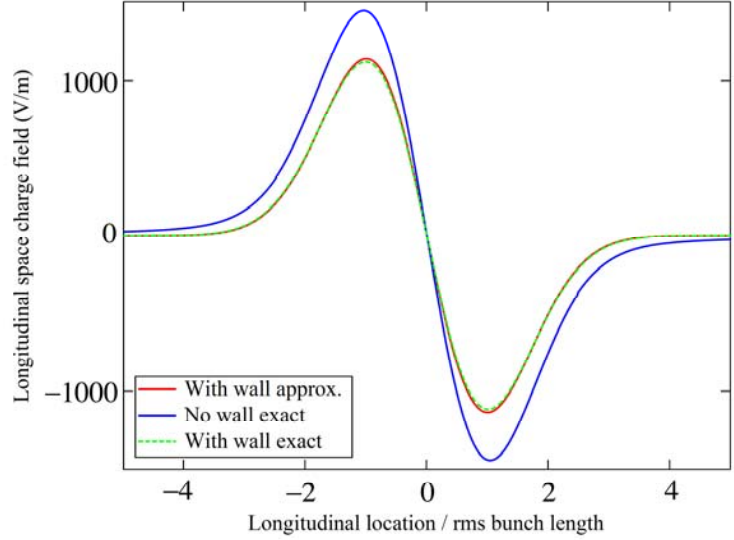

Figure 10. The longitudinal space charge field of an electron bunch in the free space (blue) and, inside a beam pipe (red and green). The blue curve is generated using eq. (51) for an electron bunch in the free space, the red curve is produced by the approximate formula, eq. (65), for a long electron bunch inside a beam pipe, and the green curve is created from the exact formula, eq. (66), for an electron bunch with arbitrary length inside a beam pipe. The parameters of the proof of CeC principle experiment are applied for all plots.

## VI. SUMMARY

In this work, we developed an analytical model to study the effect of ion shielding in the presence of a uniform electric field. The model assumes a uniform electron spatial-distribution and an anisotropic 3-D velocity distribution. We showed that the modulation in electron density, induced by a moving ion, can be expressed as a 1-D integral that depends upon both the ion's velocity and the acceleration of electrons caused by the external field. A higher modulation of the electron's peak density occurs when the acceleration of electrons is along the same direction of the ion velocity than otherwise.

We applied this model to the processes in the CeC modulator in the presence of a space charge field. Its use is valid provided that the spatial extension of the electron bunch is much larger than the corresponding Debye length. As numerical examples, we estimated influences of the longitudinal space charge field on the modulation processes in CeC used for the proof-of-principle experiment at BNL, as well as for the proposed eRHIC and LHC CeC.

For the CeC PoP experiment, our estimations show that effect is relatively mild and can reduce CeC efficiency only by a few percent. More importantly, this analysis confirmed our early estimations and conclusions that the effects of longitudinal space charge do not play a significant role in the CeC schemes proposed for the eRHIC and the LHC.

**APPENDIX A: LONGITUDINAL SPACE CHARGE FIELD OF A CHARGED BUNCH INSIDE A CONDUCTING CIRCULAR PIPE:**

For a charged particle bunch enclosed by an infinitely long conducting beam pipe with radius $b$, the electric potential, $\varphi$, at the beam pipe is equal to its value at infinity, such that the boundary condition can be set as

$$\varphi(b,\phi,z) = \varphi(r,\phi,\infty) = 0. \tag{A1}$$

We assume that the charged-particle bunch is transversely uniform with radius, $R$, and its total charge is $Q_e$. In the co-moving frame of the bunch, the Poisson equation inside the beam pipe reads

$$\nabla^2 \varphi = -\frac{Q_e}{\varepsilon_0 \pi R^2} \frac{1}{\sqrt{2\pi}\sigma_z} e^{-\frac{z^2}{2\sigma_z^2}} \theta(R-r), \tag{A2}$$

where $\theta(x)$ is the Heaviside step-function with the definition

$$\theta(x) \equiv \begin{cases} 1, & x \geq 0 \\ 0, & x < 0 \end{cases}. \tag{A3}$$

In cylindrical coordinates, eq. (A2) becomes

$$\frac{1}{r}\frac{\partial}{\partial r}\left(r\frac{\partial}{\partial r}\right)\varphi + \frac{1}{r^2}\frac{\partial^2}{\partial \phi^2}\varphi + \frac{\partial^2}{\partial z^2}\varphi = -\frac{Q_e}{\varepsilon_0 \pi R^2} \frac{1}{\sqrt{2\pi}\sigma_z} e^{-\frac{z^2}{2\sigma_z^2}} \theta(R-r). \tag{A4}$$

Since the system has cylindrical symmetry, the derivative with respect to the azimuthal angle, $\phi$, vanishes, and hence, eq. (A4) reduces to

$$\frac{1}{r}\frac{\partial}{\partial r}\left(r\frac{\partial}{\partial r}\right)\varphi + \frac{\partial^2}{\partial z^2}\varphi = -\frac{Q_e}{\varepsilon_0 \pi R^2}\frac{1}{\sqrt{2\pi}\sigma_z}e^{-\frac{z^2}{2\sigma_z^2}}\theta(R-r) . \tag{A5}$$

Taking Fourier transformation of eq. (A5) along $z$ axis yields the following:

$$\frac{1}{r}\frac{\partial}{\partial r}\left(r\frac{\partial}{\partial r}\right)\tilde{\varphi}(r,k_z) - k_z^2 \tilde{\varphi}(r,k_z) = -\frac{Q_e}{\varepsilon_0 \pi R^2}e^{-\frac{k_z^2 \sigma_z^2}{2}}\theta(R-r) , \tag{A6}$$

where,

$$\tilde{\varphi}(r,k_z) \equiv \int_{-\infty}^{\infty} e^{-ikz}\varphi(r,z)dz . \tag{A7}$$

The boundary condition for $\tilde{\varphi}(r,k_z)$ at $r = b$ is obtained from eqs. (A1) and (A7) as

$$\tilde{\varphi}(b,k_z) = 0. \tag{A8}$$

Expanding the first term of eq. (A6) yields

$$\frac{\partial^2}{\partial r^2}\tilde{\varphi}(r,k_z) + \frac{1}{r}\frac{\partial}{\partial r}\tilde{\varphi}(r,k_z) - k_z^2 \tilde{\varphi}(r,k_z) = -\frac{Q_e}{\varepsilon_0 \pi R^2}e^{-\frac{k_z^2 \sigma_z^2}{2}}\theta(R-r). \tag{A9}$$

For $r \leq R$, eq. (A9) can be rewritten into

$$\frac{\partial^2}{\partial \tilde{r}^2}\tilde{\varphi}_{in}(\tilde{r},k_z) + \frac{1}{\tilde{r}}\frac{\partial}{\partial \tilde{r}}\tilde{\varphi}_{in}(\tilde{r},k_z) - \tilde{\varphi}_{in}(\tilde{r},k_z) = -\frac{Q_e}{\varepsilon_0 \pi k_z^2 R^2}e^{-\frac{k_z^2 \sigma_z^2}{2}} , \tag{A10}$$

with

$$\tilde{r} = k_z r . \tag{A11}$$

Eq. (A10) is the inhomogeneous modified Bessel differential equation, and its solution can be expressed as

$$\tilde{\varphi}_{in}(\tilde{r},k_z) = \tilde{\varphi}_{in,h}(\tilde{r},k_z) + \tilde{\varphi}_{in,p}(\tilde{r},k_z), \tag{A12}$$

where,

$$\tilde{\varphi}_{in,h}(\tilde{r},k_z) = c_1(k_z)I_0(\tilde{r}) + c_2(k_z)K_0(\tilde{r}) \tag{A13}$$

is the solution of the homogeneous modified Bessel differential equation, i.e.,

$$\frac{\partial^2}{\partial \tilde{r}^2}\tilde{\varphi}_{in,h}(\tilde{r},k_z) + \frac{1}{\tilde{r}}\frac{\partial}{\partial \tilde{r}}\tilde{\varphi}_{in,h}(\tilde{r},k_z) - \tilde{\varphi}_{in,h}(\tilde{r},k_z) = 0 , \tag{A14}$$

and $\tilde{\varphi}_{in,p}(k_z,\tilde{r})$ is the particular solution satisfying eq. (A10). Since the driving term in eq. (A10) is independent of $\tilde{r}$, the particular solution reads

$$\tilde{\varphi}_{in,p}(\tilde{r},k_z) = \frac{Q_e}{\varepsilon_0 \pi k_z^2 R^2}e^{-\frac{k_z^2 \sigma_z^2}{2}} . \tag{A15}$$

Inserting eqs. (A13) and (A15) into eq. (A12) yields

$$\tilde{\varphi}_{in}(\tilde{r},k_z) = c_1(k_z)I_0(\tilde{r}) + c_2(k_z)K_0(\tilde{r}) + \frac{Q_e}{\varepsilon_0 \pi k_z^2 R^2}e^{-\frac{k_z^2 \sigma_z^2}{2}} . \tag{A16}$$

Requiring $\tilde{\varphi}_{in}(\tilde{r}=0)$ to being finite leads to

$$\tilde{\varphi}_{in}(\tilde{r},k_z) = c_1(k_z) I_0(\tilde{r}) + \frac{Q_e}{\varepsilon_0 \pi k_z^2 R^2} e^{-\frac{k_z^2 \sigma_z^2}{2}}. \tag{A17}$$

For $r > R$, eq. (A9) becomes the homogeneous modified Bessel differential equation

$$\frac{\partial^2}{\partial \tilde{r}^2} \tilde{\varphi}_{out}(\tilde{r},k_z) + \frac{1}{\tilde{r}} \frac{\partial}{\partial \tilde{r}} \tilde{\varphi}_{out}(\tilde{r},k_z) - \tilde{\varphi}_{out}(\tilde{r},k_z) = 0, \tag{A18}$$

that has solutions of the form:

$$\tilde{\varphi}_{out}(\tilde{r},k_z) = d_1(k_z) I_0(\tilde{r}) + d_2(k_z) K_0(\tilde{r}). \tag{A19}$$

Applying the boundary condition, eq. (A8), leads to

$$d_2(k_z) = -d_1(k_z) \frac{I_0(k_z b)}{K_0(k_z b)}, \tag{A20}$$

and hence, eq. (A19) becomes

$$\tilde{\varphi}_{out}(\tilde{r},k_z) = d_1(k_z) \left[ I_0(\tilde{r}) - \frac{I_0(k_z b)}{K_0(k_z b)} K_0(\tilde{r}) \right]. \tag{A21}$$

The two remaining coefficients, $c_1(k_z)$ and $d_1(k_z)$, are determined by the conditions at the beam's boundary, $r = R$, which read

$$\tilde{\varphi}_{out}(k_z R, k_z) = \tilde{\varphi}_{in}(k_z R, k_z), \tag{A22}$$

and,

$$\left. \frac{\partial}{\partial \tilde{r}} \tilde{\varphi}_{out}(\tilde{r},k_z) \right|_{\tilde{r}=k_z R;\ k_z=const.} = \left. \frac{\partial}{\partial \tilde{r}} \tilde{\varphi}_{in}(\tilde{r},k_z) \right|_{\tilde{r}=k_z R;\ k_z=const.}. \tag{A23}$$

Eqs. (A22) and (A23) produce

$$d_1(k_z) = -\frac{Q_e}{\varepsilon_0 \pi k_z R^2} \frac{R I_1(k_z R) K_0(k_z b)}{I_0(k_z b)} e^{-\frac{k_z^2 \sigma_z^2}{2}}, \tag{A24}$$

and,

$$c_1(k_z) = -\frac{Q_e}{\varepsilon_0 \pi k_z R} e^{-\frac{k_z^2 \sigma_z^2}{2}} \left[ I_1(k_z R) \frac{K_0(k_z b)}{I_0(k_z b)} + K_1(k_z R) \right], \tag{A25}$$

where we used the relations

$$\frac{d}{dz} K_0(z) = -K_1(z), \tag{A26}$$

and,

$$\frac{d}{dz} I_0(z) = I_1(z), \tag{A27}$$

and.

$$I_\nu(z) K_{\nu+1}(z) + I_{\nu+1}(z) K_\nu(z) = \frac{1}{z}. \tag{A28}$$

Inserting eqs. (A24) and (A25) into eqs. (A21) and (A17) generates

$$\tilde{\varphi}_{in}(r,k_z) = -\frac{Q_e}{\varepsilon_0 \pi R^2} e^{-\frac{k_z^2 \sigma_z^2}{2}} \left\{ \frac{R}{k_z} \left[ \frac{I_1(k_z R) K_0(k_z b) + K_1(k_z R) I_0(k_z b)}{I_0(k_z b)} \right] I_0(k_z r) - \frac{1}{k_z^2} \right\}, \tag{A29}$$

for $r \leq R$, and,

$$\tilde{\varphi}_{out}(r,k_z) = -\frac{Q_e R}{\varepsilon_0 \pi k_z R^2} e^{-\frac{k_z^2 \sigma_z^2}{2}} \frac{I_1(k_z R)}{I_0(k_z b)} \left[ K_0(k_z b) I_0(k_z r) - I_0(k_z b) K_0(k_z r) \right], \quad (A30)$$

for $R < r \leq b$.

The electric potential inside the bunch is given by the inverse Fourier transformation of eq. (A29), i.e.,

$$\varphi_{in}(r,z) = \frac{1}{2\pi} \int_{-\infty}^{\infty} \tilde{\varphi}_{in}(r,k_z) e^{ik_z z} dk_z, \quad (A31)$$

That leads to the longitudinal electric field:

$$E_{z,in}(r,z) = -\frac{\partial}{\partial z} \varphi_{in}(r,z) = -\frac{i}{2\pi} \int_{-\infty}^{\infty} k_z \tilde{\varphi}_{in}(r,k_z) e^{ik_z z} dk_z. \quad (A32)$$

On the bunch axis, $r = 0$, and the longitudinal electric field is

$$\begin{aligned} E_{z,in}(0,z) &= -\frac{i}{2\pi} \int_{-\infty}^{\infty} k_z \tilde{\varphi}_{in}(0,k_z) e^{ik_z z} dk_z \\ &= \frac{Q_e}{\varepsilon_0 \pi R} \frac{i}{2\pi} \int_{-\infty}^{\infty} e^{-\frac{k_z^2 \sigma_z^2}{2}} \left\{ \frac{I_1(k_z R) K_0(k_z b)}{I_0(k_z b)} + K_1(k_z R) - \frac{1}{k_z R} \right\} e^{ik_z z} dk_z, \quad (A33) \\ &= \frac{Q_e}{\varepsilon_0 \pi R} \frac{i}{2\pi} \int_{0}^{\infty} \left[ f(k_z) e^{ik_z z} + f(-k_z) e^{-ik_z z} \right] dk_z \end{aligned}$$

with

$$f(k_z) = e^{-\frac{k_z^2 \sigma_z^2}{2}} \left\{ \frac{I_1(k_z R) K_0(k_z b) + I_0(k_z b) K_1(k_z R)}{I_0(k_z b)} - \frac{1}{k_z R} \right\}. \quad (A34)$$

The function, $f(k_z)$, is an odd function of $k_z$. To prove it, it is sufficient to show that the function,

$$h(k_z) = I_1(k_z R) K_0(k_z b) + I_0(k_z b) K_1(k_z R) \quad (A35)$$

is odd, or, more explicitly,

$$h(k_z) + h(-k_z) = I_1(k_z R)\left[ K_0(k_z b) - K_0(-k_z b) \right] + I_0(k_z b)\left[ K_1(k_z R) + K_1(-k_z R) \right], \quad (A36)$$

vanishes for any real value of $k_z$. The integral representations of the modified Bessel function of the 0$^{th}$ order read [10]

$$K_0(z) = -\frac{1}{\pi} \int_0^{\pi} e^{\pm z \cos\theta} \left[ \gamma + \ln(2z \sin^2 \theta) \right] d\theta, \quad (A37)$$

and,

$$I_0(z) = \frac{1}{\pi} \int_0^{\pi} e^{\pm z \cos\theta} d\theta, \quad (A38)$$

where,

$$\gamma \equiv \lim_{n \to \infty} \left( 1 + \frac{1}{2} + \frac{1}{3} + \cdots + \frac{1}{n} - \ln n \right) \approx 0.5772156649... \quad (A39)$$

is the Euler's constant. It follows from eq. (A37) that

$$K_0(z) - K_0(-z) = -\frac{1}{\pi}\int_0^\pi e^{\pm z\cos\theta}\left[\ln(2z\sin^2\theta) - \ln(-2z\sin^2\theta)\right]d\theta$$

$$= \frac{\ln(-1)}{\pi}\int_0^\pi e^{\pm z\cos\theta}d\theta \qquad , \qquad (A40)$$

$$= i\pi(2n+1)I_0(z)$$

with $n$ being an arbitrary integer. Taking the first derivative of eq. (A40) gives

$$K_1(z) + K_1(-z) = -i\pi(2n+1)I_1(z). \tag{A41}$$

Using eqs. (A40) and (A41), eq. (A36) becomes

$$h(k_z) + h(-k_z) = i\pi(2n+1)I_1(k_z R)I_0(k_z b) - i\pi(2n+1)I_1(k_z R)I_0(k_z b) = 0, \tag{A42}$$

and consequently, we proved that $f(k_z)$ is an odd function, i.e.,

$$f(-k_z) = -f(k_z). \tag{A43}$$

Inserting eq. (A43) into eq. (A33), we obtain

$$E_{z,in}(0,z) = -\frac{Q_e}{\varepsilon_0 \pi^2 R}\int_0^\infty f(k_z)\sin(k_z z)dk_z$$

$$= -\frac{Q_e}{\varepsilon_0 \pi^2 R}\int_0^\infty e^{-\frac{k_z^2\sigma_z^2}{2}}\left\{\frac{I_1(k_z R)K_0(k_z b)}{I_0(k_z b)} + K_1(k_z R) - \frac{1}{k_z R}\right\}\sin(k_z z)dk_z \qquad . \qquad (A44)$$

It often is convenient to express eq. (A44) in terms of the normalized variables

$$E_{z,in}(0,\bar{z}) = -\frac{Q_e}{\varepsilon_0 \pi^2 \bar{R}\sigma_z^2}\int_0^\infty e^{-\frac{\xi^2}{2}}\left\{\frac{I_1(\xi\cdot\bar{R})K_0(\xi\cdot\bar{b})}{I_0(\xi\cdot\bar{b})} + K_1(\xi\cdot\bar{R}) - \frac{1}{\xi\bar{R}}\right\}\sin(\xi\cdot\bar{z})d\xi, \tag{A45}$$

with $\bar{R} = R/\sigma_z$, $\bar{z} = z/\sigma_z$ and $\bar{b} = b/\sigma_z$. Applying the asymptotic behavior of the modified Bessel function at $k_z \to 0$ yields

$$\lim_{k_z\to 0}\left[f(k_z)\sin(k_z z)\right] = \lim_{k_z\to 0}\left\{\left[-\frac{k_z R}{2}\ln(k_z b) + \frac{1}{k_z R} - \frac{1}{k_z R}\right]\sin(k_z z)\right\} = 0. \tag{A46}$$

At $k_z \to \infty$, the asymptotic behaviors of the modified Bessel functions are

$$I_{0,1}(x) \sim \frac{e^x}{\sqrt{2\pi x}},$$

$$K_{0,1}(x) \sim \sqrt{\frac{\pi}{2x}}e^{-x},$$

and hence, it follows that

$$\lim_{k_z\to\infty}\left[f(k_z)\sin(k_z z)\right] = \lim_{k_z\to\infty}\left\{e^{-\frac{k_z^2\sigma_z^2}{2}}\left(\sqrt{\frac{\pi}{2k_z R}}e^{-k_z(2b-R)} + \sqrt{\frac{\pi}{2k_z R}}e^{-k_z R} - \frac{1}{k_z R}\right)\sin(k_z z)\right\} = 0$$

. (A47)

Eqs. (A46) and (A47) suggest that the integrand of the integral in eq. (A44), is a finite real function. Fig. A1 compares the results in eq. (A44) with the previously derived longitudinal space charge field in the absence of the beam pipe and with that in the presence of the beam pipe but with long beam approximation. As shown in Fig. A1(a), when the beam frame's bunch length, 1.26 cm, is much smaller than the beam-pipe's radius, the exact solution (green) in eq. (A44) overlaps with that for an open beam (blue), i.e., without a beam pipe. Fig. A1(b) shows that as the beam frame's bunch length increases to 1.26 m, the exact solution overlaps with that of the long-bunch approximation (red). With a R.M.S. bunch length of 5cm, Fig. A1(c) shows that the exact solution deviates from that in both of the other solutions.

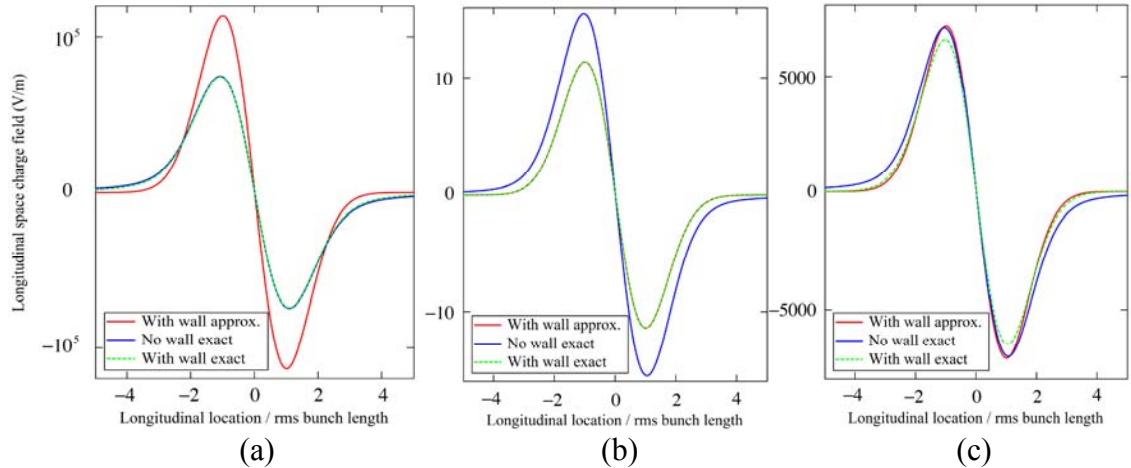

(a)  (b)  (c)

Figure A1: The exact longitudinal space charge field in the presence of the 5cm-radius beam pipe. The abscissas are the longitudinal locations in units of the R.M.S. bunch length and the ordinates are the longitudinal space charge field in units of V/m. The green dashed curve shows the field calculated from the exact solution, i.e., eq. (A44), the blue solid curve shows the field from the electron bunch, without considering the beam pipe , and the red solid curve shows the field calculated from an approximate formula where shielding from the beam pipe is considered but the electron bunch is assumed to be much longer than the beam pipe's radius. (a) Calculated for PoP parameters but with 1.26 cm beam frame R.M.S. bunch length; (b) calculated with 1.26 m beam frame R.M.S. bunch length; and, (c) calculated with 5 cm beam frame R.M.S. bunch length.